\input harvmac
\overfullrule=0pt
\newcount\figno
\figno=0
\def\fig#1#2#3{
\par\begingroup\parindent=0pt\leftskip=1cm\rightskip=1cm
\parindent=0pt
\baselineskip=11pt
\global\advance\figno by 1
\midinsert
\epsfxsize=#3
\centerline{\epsfbox{#2}}
\vskip 12pt {\bf Fig. \the\figno:} #1\par
\endinsert\endgroup\par }
\def\figlabel#1{\xdef#1{\the\figno}}
\def\encadremath#1{\vbox{\hrule\hbox{\vrule\kern8pt\vbox{\kern8pt
\hbox{$\displaystyle #1$}\kern8pt}
\kern8pt\vrule}\hrule}}

\Title{\vbox{\baselineskip12pt
\hbox{hep-th/0101118}
\hbox{OHSTPY-HEP-T-00-033} }} {\vbox{\centerline{Gravity on $AdS_3$
and flat connections}\centerline{ in the boundary CFT }}}
\smallskip
\bigskip
\centerline{Samir D. Mathur}
\smallskip
\centerline{{\it Department of Physics, The Ohio State University}}
\centerline{\it Columbus, OH. 43210,  U.S.A.}
\bigskip

\medskip
\bigskip
\bigskip

\noindent

\def\p{\partial}

We consider the CFT that arises from the D1-D5 system in the presence of a
constant background gauge potential which couples to the R-charge of 
the theory:
this potential effectively changes the periodicities of the fermions. 
By the  AdS/CFT
correspondence, the effect of this connnection should be obtained by finding a
smooth solution to the vector field in AdS space which couples to the constant
mode of the R-current in the CFT. We investigate such solutions for 
small values of the
connection, and contrast these with spacetimes which have `Wilson
lines for the bulk ' --  spacetimes that are locally AdS times a 
sphere but have a global
deformation. The latter class are in general singular spacetimes. We comment on
some aspects of the recently found geometries corresponding to the D1-D5 state
with angular momentum, observing relations between scales in the microscopic
theory and scales in  the geometry.

\Date{January, 2001}

\newsec{Introduction}

The D1-D5 system has been crucial in the study of black holes in string theory,
since it provides a microscopic description of black hole entropy and Hawking
radiation\ref\blackhole{A.~Strominger and C.~Vafa, ``Microscopic Origin of the
Bekenstein-Hawking Entropy,'' Phys.\ Lett.\  {\bf B379}, 99 (1996)
[hep-th/9601029], C.~G.~Callan and J.~M.~Maldacena, ``D-brane Approach to Black
Hole Quantum Mechanics,'' Nucl.\ Phys.\  {\bf B472}, 591 (1996)
[hep-th/9602043], S.~R.~Das and S.~D.~Mathur, ``Comparing decay rates for black
holes and D-branes,'' Nucl.\ Phys.\  {\bf B478}, 561 (1996) [hep-th/9606185],
S.~R.~Das and S.~D.~Mathur, ``Interactions involving D-branes,'' 
Nucl.\ Phys.\  {\bf
B482}, 153 (1996) [hep-th/9607149], A.~Dhar, G.~Mandal and S.~R.~Wadia,
``Absorption vs decay of black holes in string theory and T-symmetry,'' Phys.\
Lett.\  {\bf B388}, 51 (1996) [hep-th/9605234], J.~Maldacena and A.~Strominger,
``Black hole greybody factors and D-brane spectroscopy,'' Phys.\ 
Rev.\  {\bf D55},
861 (1997) [hep-th/9609026].}.  According to the conjecture by Maldacena
\ref\maldacena{J.~Maldacena, ``The large N limit of superconformal 
field theories
and supergravity,'' Adv.\ Theor.\ Math.\ Phys.\  {\bf 2}, 231 (1998)
[hep-th/9711200].},  The D1-D5 CFT is dual to IIB string theory compactified on
$AdS_3\times S^3\times M_4$ where the 4-manifold $M_4$ can be $T^4$ or $K3$.
When this duality is examined in more detail, one encounters the 
question of how
the boundary CFT is related to the bulk spacetime \ref\gkpw{S.~S.~Gubser,
I.~R.~Klebanov and A.~M.~Polyakov, ``Gauge theory correlators from non-critical
string theory,'' Phys.\ Lett.\  {\bf B428}, 105 (1998)
[hep-th/9802109]}\ref\wittenone{ E.~Witten, ``Anti-de Sitter space and
holography,'' Adv.\ Theor.\ Math.\ Phys.\  {\bf 2}, 253 (1998)
[hep-th/9802150].}.  For the Euclidean theory the CFT is placed at the boundary
of a large ball shaped region of $AdS$ space. For
  Lorentzian signature,  the `global
$AdS$' space is dual to a CFT  on a cylindrical boundary with the state on the
cylinder being the Neveu-Schwarz (NS) vaccum. The near horizon geometry of
branes, on the other hand, gives  the Poincare patch of $AdS$ space.

The question that naturally arises is whether one can find a 
representation of the
Ramond (R) vaccum of the CFT. The  zero mass BTZ black hole solution of the
gravity theory is expected to be a Ramond sector ground state
\ref\henn{O.~Coussaert and M.~Henneaux,
Phys.\ Rev.\ Lett.\  {\bf 72}, 183 (1994)
[hep-th/9310194].}\ref\ads{J.~Maldacena and A.~Strominger, ``AdS(3)
black holes and a stringy exclusion principle,'' JHEP {\bf 9812}, 005 (1998)
[hep-th/9804085]}\ref\deboer{ J.~de Boer, ``Large N Elliptic Genus and AdS/CFT
Correspondence,'' JHEP {\bf 9905}, 017 (1999) [hep-th/9812240].}. 
The Ramond sector has
periodic fermions, while the NS sector has anti-periodic fermions. 
One would therefore like to
ask if there are supergravity backgrounds with all other possible 
intermediate boundary
conditions on the fermions.  In particular, we can  consider the CFT 
in the presence
of a constant background gauge potential coupling to the R-charge of 
the CFT. This
R charge is carried by the fermions of the theory, and such a gauge potential
effectively modifies the boundary conditions on the fermions.  The family of
theories with different values of the gauge potential can be related 
by a `spectral
flow'. Thus  we can start with the NS vaccum (at zero gauge 
potential)   and look at
the states that are obtained as we continuounsly increase the potential.

The R-symmetry of the CFT is $su(2)\oplus su(2)$, with the first $su(2)$ being
carried by the left movers and the second $su(2)$ being carried by the right
movers.  Let the background gauge field  be called ${\bf 
A}^{ext}_{CFT}$; it will have
the parts $ A^{ext,a}_L$ and $A^{ext,a}_R$  ($a=1,2,3)$ for the left and right
movers. Let the currents of the CFT be ${\bf J}$; these will have the 
left and right
parts
$J_L^a$ and $J_R^a$.  At least to leading order in the potential the 
CFT will have an
action
\eqn\gone{S=S_0+{\bf A}^{ext}_{CFT}{\bf J}} where $S_0$ is the action in the
absence of any background potential. This minimal coupling is what we would get
for example by looking at the orbifold CFT that we get at the 
orbifold point of the
D1-D5 system, and we expect that it will be valid more generally when the
current ${\bf J}$ has been appropriately defined in the presence of 
the gauge field.

The external gauge fields ${\bf A}^{ext}_{CFT}$ get related to the 
fields  ${\bf
A}_{SUGRA}$ of the supergravity theory which are dual to the CFT currents ${\bf
J}$, as follows.  If $\phi_i$ are fields of the supergravity theory 
dual to operators $
O_i$ in the CFT, then we compute correlators of the $O_i$ from the relation
\eqn\bone{\int DX e^{-S_{CFT}- O_i\phi_i^b}=e^{-S_{SUGRA}(\phi_i^b)}} where the
supergravity action is computed on the field configuration obtained  by
extremising with the $\phi_i$ fixed to have the values  $\phi_i^b$ at 
the boundary
of $AdS$.  If we want to turn on an external gauge field in the CFT, 
then we would
get
\eqn\btwo{\eqalign{\int DX &e^{-S_{CFT}-O_i\phi_i- {\bf 
A}^{ext}_{CFT}{\bf J}}=\int
DXe^{-S_{CFT}-O_i\phi_i}[1- {\bf A}^{ext}_{CFT}{\bf J}+\dots]\cr
&=e^{-S_{SUGRA}[\phi_i^b]}+{\bf A}^{ext}_{CFT}[{\delta\over \delta {\bf
A}_{SUGRA}^b} e^{-S_{SUGRA}[\phi_i^b+{\bf A}^b_{SUGRA}]}]_{{\bf
A}_{SUGRA}^b=0}+\dots\cr &=e^{-S_{SUGRA}[\phi^b_i, {\bf A}_{SUGRA}^b={\bf
A}^{ext}_{CFT}]}\cr }} at least to leading order in the gauge field 
where the issue of
contact interactions between multiple ${\bf J}$ insertions does not 
arise; we expect
that with a suitable definition of the field ${\bf A}_{SUGRA}$ any 
magnitide of ${\bf
A}^{ext}_{CFT}$ will be identified with a value of ${\bf A}^b_{SUGRA}$.

Thus it would appear that to take into account a constant external 
gauge potential
in the boundary CFT we must look for smooth solutions to the supergravity
equations that take a specified value at the boundary  for the gauge 
field ${\bf
A}_{SUGRA}$  that is dual to the current operator ${\bf J}$ in the CFT; we then
evaluate all other correlation functions in the presence of this 
supergravity field.

In an interesting paper
\ref\wadia{J.~R.~David, G.~Mandal, S.~Vaidya and S.~R.~Wadia, ``Point mass
geometries, spectral flow and AdS(3)-CFT(2) correspondence,'' Nucl.\ 
Phys.\  {\bf
B564}, 128 (2000) [hep-th/9906112].} it was argued (extending earlier 
results of
\ref\town{J.~M.~Izquierdo and P.~K.~Townsend, ``Supersymmetric space-times in
(2+1) adS supergravity models,'' Class.\ Quant.\ Grav.\  {\bf 12}, 895 (1995)
[gr-qc/9501018].}) that a family of $AdS_3$ geometries with varying conical
defect angles would be the gravity descriptions of the family of CFT states
connected by spectral flow between the NS and R sectors. These 
spacetimes had in
addition to the conical defect a Wilson line of a gauge field from 
the supergravity
multiplet: this flat connection would change the boundary condition 
for the fields
charged under the internal $su(2)\oplus su(2)$ symmetry group of the CFT.

In  \wadia\   the supergravity theory was  extended supergravity in three
dimensions, and the gauge fields thus arose as superpartners of the 
gravity fields.
But if the $AdS_3$ spacetime is obtained by compactification of 10-d IIB string
theory, then it must be possible to identify this gauge field among 
the fields of the
10-d supergravity multiplet, and look at the 10-d spacetime 
corresponding to the
spacetimes of interest.   We carry out such an identification, and 
find that the IIB
spacetimes that represent the $AdS$ construction of \wadia\ appear to be
spacetimes that are locally $AdS_3\times S^3$ (we will suppress the compact
manifold $M_4$ in most of what follows) but which have 
identifications that make
them not globally a simple product of $AdS_3$ and $S^3$. But these 
spacetimes are
generally {\it singular} at $r=0$, and thus would not arise as smooth 
solutions of
the supergravity fields with given values at the boundary.  We
comment on this issue again at the end of this paper.

In this paper we do the following:
\bigskip (a) \quad  We look at the fields in supergravity that carry 
the quantum
numbers of the operators ${\bf J}$ in the CFT. We solve their coupled 
equations,
and observe the `chiral' nature of their solutions: The behavior of 
the field on
$S^3$ must correlate with the behavior in $AdS^3$. This is the case because the
former derermines which $su(2)$ out of the $su(2)\oplus su(2)$  R symmetry the
field carries, while the behavior of the field near the  $AdS_3$ boundary
determines whether it couples to left or right movers in the CFT.
\bigskip (b)\quad We examine the spacetimes that are locally of the form
$AdS_3\times S^3$ but are not globally a direct product. We look at 
three different
kinds of deformations of $AdS_3\times S^3$ in this category. We may term these
deformations `Wilson lines'  of the bulk supergravity theory.  (By 
what was said
above these may not represent Wilson lines in the boundary CFT.) These deformed
spacetimes are in general singular spacetimes (except when the `twist' equals a
multiple of
$2\pi$; then it can be removed by a coordinate transformation).  We look at the
wavefunctions and energy levels for localized states of a scalar field in the
backgrounds that have `twists' and conical defects (which are two of the above
mentioned three kinds of deformations). A subclass of these wavefunctions
generalise the wavefunctions that were constructed in \ads\  as 
duals to primary
fields  of the CFT  when the spacetime was  globally  $AdS_3\times 
S^3$.  However,
the significance of these geometries (and these localised 
wavefunctions)  in the
context of  the AdS/CFT correspondence is not quite clear.
\bigskip (c)\quad In some recent papers \ref\bal{V.~Balasubramanian, 
J.~de Boer,
E.~Keski-Vakkuri and S.~F.~Ross, ``Supersymmetric conical defects: Towards a
string theoretic description  of black hole formation,''
hep-th/0011217.}\ref\mm{J.~Maldacena and L.~Maoz, ``De-singularization by
rotation,'' hep-th/0012025.} a set of geometries was given that corresponded to
the D1-D5 system with no momentum charge but with nonzero angular
momentum.  We  comment on several interesting aspects of these geometries. In
particular we note that from the microscopic viewpoint these systems are in a
different `phase' from the D1-D5-momentum black holes.  We observe several
interesting correspondences between scales of the gravity theory and 
scales of the
microscopic dual.

\bigskip

The plan of this paper is the following. In section 2 we look at the linearized
equations of the fields that carry the quantum numbers of the current operators
in the CFT, and analyze some of their solutions. In section 3 we look 
at `pure gauge'
solutions that are spacetimes that are locally but not globally 
$AdS_3\times S^3$;
these are finite deformations of the spacetime  that appear at the 
linearised level
as pure gauge deformations of the fields. In section 4 we look at 
solutions to the
scalar wave equation around some of the deformed backgrounds  studied in
section 3. In section 5 we examine some interesting relations between the
solutions presented in \bal\mm\  and  the microscopic theory of the D1-D5
system.  Section 6 is a summary and discussion.

\bigskip

{\bf Note added:}\quad Several of the  computations described in this paper
were performed at various points
over the past year, as part of a general study of the AdS/CFT 
correspondence for
the D1-D5 system. These notes have been brought to the form of the present
paper partly due to the interest in this issue generated by recent 
studies of the
spacetimes corresponding to states in the Ramond sector of the CFT
\bal\mm . We discuss the relation of our computations to these latter studies
towards the end of this paper,  and comment on some interesting aspects of the
ideas emerging from the above references.

\newsec{The linearised theory}

\subsec{Field equations}

Let us consider the compactification of type IIB string theory on a 
K3 or $T^4$. We
will get a set of low energy supergravity fields. We will focus on 
the graviton and
the 2-forms arising in the 6 noncompact directions. The lagrangian 
and linearised
equations of these fields can be read off from
\ref\deger{S.~Deger, A.~Kaya, E.~Sezgin and P.~Sundell, ``Spectrum of 
D = 6, N = 4b
supergravity on AdS(3) x S(3),'' Nucl.\ Phys.\  {\bf B536}, 110 (1998)
[hep-th/9804166].} where a general $D=6, N=4b$ supergravity theory was
studied. (If the compact space is a $K3$ then there will be no 
1-forms from in 6-d;
for $T^4$ there will be 1-forms but they are not the ones of interest 
to us below.)
In subsection 2.3  below we will obtain thes equations directly from IIB
supergravity in 10-D, reduced to 6-D on $T^4$.

  The 2-form fields can be split into self-dual and anti-self-dual 
sets, and one of the
self-dual fields acquires a vacuum expectation value to give a 
compactification  the
6-d space to a geometry $AdS_3\times S^3$.  We will use indices $A,B, \dots$ to
represent 6-d indices,
$\mu,\nu,\dots$ to represent $AdS_3$ indices, and $a,b, \dots$ to 
represent $S^3$
indices. We call the $AdS$ coordinates $x$ and the $S^3$ coordinates 
$y$. We will
also ignore the scalars in the discussion below, and so will not 
distinguish between
the elementary field strengths $G=dB$ and the modified field strengths
$H$. Letting  the radii of the two factors be unity each, we get the 
curvature tensor
and self dual field stengths
\eqn\eone{R_{\mu\nu\rho\sigma}=-(g_{\mu\rho}g_{\nu\sigma}-g_{\nu\rho}
g_{\mu\sigma}),
~~R_{abcd}=g_{ac}g_{bd}-g_{ad}g_{bc}}
\eqn\etwo{H_{\mu\nu\rho}=\partial_\mu B_{\nu\rho}+\partial_\nu
B_{\rho\mu}+\partial_\rho B_{\mu\nu}, ~~~H_{abc}=\partial_a B_{bc}+\partial_b
B_{ca}+\partial_c B_{ab} }
\eqn\ethree{H_{\mu\nu\rho}=\epsilon_{\mu\nu\rho}, ~~~H_{abc}=\epsilon_{abc}}

The sphere $S^3$ has the symmetry group $SO(4)=(SU(2)\times SU(2))/Z_2$, and
thus the algebra of diffeomorphisms is $su(2)\oplus su(2)$. We identify these
$su(2)$ factors with the left and right current algebras of the 
boundary CFT.  To
find the fields dual to the currents $J_L^i, J_R^i$ of the boundary 
theory we must
look at supergravity fields that are vectors in the $AdS_3$ and vectors on the
sphere $S^3$. The components of the graviton $h^a_\mu$ are such an object; we
also have $B^a{}_\mu$ from the 2-form field which acquires an 
expectation value.
The other 2-form fields in the theory are related  by symmetry: the 
self dual fields
had an $SO(5)$ symmetry which is broken to $SO(4)$ when we single out one field
to give it an expectation value, and the anti-self-dual fields have a $SO(n)$
symmetry, where $n$ is the number of tensor multiplets in the $D=6, N=4b$
theory.  Thus we do not expect fields from these categories to be dual to the
currents of the boundary theory (the currents do not carry these
$SO(4)$ or $SO(n)$ symmetries.)

In appendix  A we describe explicitly the vector fields that correspond to the
symmetries of the sphere, and indicate their separation into the two
$su(2)$ factors. But in the boundary CFT we must find the left movers 
to have one
such $su(2)$ and the right movers to have the other one. Thus
  $h^a_\mu$ or $B^a{}_\mu$ must yield a correlation between the behavior of the
index of the sphere and the index in the $AdS$. Such a correlation 
arises because
of Chern-Simmons like couplings between the $h^a_\mu$ and the $B^a{}_\mu$
fields, and in this section we would like to analyse the dynamics of 
these coupled
fields at the linearized level.

Before we write the explicit equations, let us recall why the two 
fields $h^a_\mu$,
which is the perturbation of the graviton,  and $b_{a\mu}$, which is 
a perturbation
of the 2-form field $B$,  are coupled at the linearized level. The 
action has a term
$H_{ABC}H^{ABC}$. Suppose we have an excitation
$b_{a\mu}$. If we have  derivatives in the directions
$\mu, \nu$ then this gives a component
$\delta H_{\nu a\mu}=\partial_\nu b_{a\mu}-\partial_\mu b_{a\nu}$.  This field
strength couples to the background field strength $H_{\mu\nu\lambda}$  if there
is a nonzero value fo $h_{\mu a}$, thus giving a quadratic order 
contribution to
$H^2$ involving one $B_{a\mu}$ and one $h_{\mu a}$ perturbation.
  This mechanism is similar to that which mixes the two 2-form fields in the
presence of a background 5-form field strength in $AdS_5\times S^5$
\ref\van{H.~J.~Kim, L.~J.~Romans and P.~van Nieuwenhuizen, ``The Mass Spectrum
Of Chiral N=2 D = 10 Supergravity On S**5,'' Phys.\ Rev.\  {\bf D32}, 
389 (1985).}

Following \deger\ we write
\eqn\rone{h_{\mu a}=K_{\mu a}, ~~~b_{\mu a}=Z_{\mu a } }
\eqn\rtwo{K_{\mu a }=K_\mu^{1,\pm 1}(x)Y_a^{1,\pm 1}(y), ~~~Z_{\mu
a}=Z_\mu^{1,\pm 1}(x)Y_a^{1,\pm 1}(y) }

The equations relating $K$ and $Z$ are \deger\
\eqn\rfour{K_{\mu;\nu}^{1, \pm 1}{}^{;\nu}- K^{1, \pm 1}_\nu 
{}^{;\nu}{}_{;\mu}-2
K_\mu^{1, \pm 1} +4\epsilon_{\mu}{}^{\nu\lambda}\partial_{\nu}Z_\lambda^{1,
\pm 1}\mp 8Z_\mu^{1, \pm 1}=0}
\eqn\rfive{\epsilon_{\mu}{}^{\nu\lambda}\partial_{\nu}Z_{\lambda}^{1, \pm
1}\pm2Z_\mu^{1, \pm 1}+K_\mu^{1, \pm 1}=0} Let us work with the upper sign.
Writing
\eqn\rseven{A_\mu\equiv K_\mu^{1, 1}, ~~~B_\mu=2 Z_\mu ^{1,1} }
\eqn\rthree{F_{\mu\nu}=\partial_\mu A_\nu-\partial_\nu A_\mu,
~~~G_{\mu\nu}=\partial_\mu A_\nu-\partial_\nu A_\mu } we get the equations
\eqn\reight{F^{\nu\mu}{}_{;\nu}+2\epsilon^{\mu\nu\rho}G_{\nu\rho}=0}
\eqn\rnine{\epsilon^{\mu\nu\rho}G_{\nu\rho}+4(A^\mu+B^\mu)=0}

Let us write
\eqn\rten{\eqalign{F_{\mu\nu}=(dA)_{\mu\nu}&=A_{\nu, \mu}-A_{\mu, \nu},
~~(*F)_\mu={1\over 2}\epsilon_{\mu\nu\rho}F^{\nu\rho},\cr
(d*F)_{\mu\nu}&=\partial_\mu(*F)_\nu-\partial_\nu(*F)_\mu,
~~~(*d*F)_\mu={1\over 2}\epsilon_{\mu\nu\rho}(d*F)^{\nu\rho}\cr}} Then the
equation \reight\ becomes
\eqn\rel{(*d*F)+4(*dB)_\mu=0} This gives
\eqn\rtw{d*F+4B=0, ~~~B=-{1\over 4} F +d\Lambda} The equation \rnine\ becomes
\eqn\rthir{2*G+4(A+B)=0} Thus once we find a solution to \rtw\ with
$\Lambda=0$, adding in the term $d\Lambda$ to $B$ does not change $G=dB$, and
from \rthir\ requires
$A\rightarrow A-d\Lambda$. This freedom
\eqn\rfourt{B\rightarrow B+d\Lambda, ~~A\rightarrow A-d\Lambda} gives the
pure gauge freedom of the equations; we will set $\Lambda=0$ in the calculation
below and return to the issue of flat connections later. Then $*G=-2A+{1\over
2}*F$, and \rel\  becomes
\eqn\rfift{*d*dA-8A+2*dA=0} We can write this as
\eqn\rsixt{(*d+4)(*d-2)A=0} which can be solved in two ways
\eqn\rsevent{(*d+4)A=0 ~~~{\rm or}~~~ (*d-2)A=0} If on the other hand we had
chosen the lower sign in passing from \rfour , \rfive\ to \reight , 
\rnine\ then we
would have a replacement
$*\rightarrow -*$, and the equations would be different. Thus the different
1-forms on the sphere $S^3$ lead to different solutions in the 
$AdS_3$ spacetime.

\subsec{A special solution}

We write in explicit form the equation \rsevent\ in Appendix D, and 
reduce it to
an equation of hyoergeometric form.  Here  we extract only the 
solutions relevant
to the constant connection in the boundary CFT.  Let us solve for the condition
$(*d-2)A=0$. In Appendix D we have considered solutions starting with the form
$$A_u=e^{-i\alpha u} f(r), ~~~A_r=e^{-i\alpha u} h(r), 
~~~A_v=e^{-i\alpha u} q(r)$$
where $u=t+\phi, v=t-\phi$.  To get a constant connection at the 
boundary we set
$\alpha=0$.  Then using the relations in Appendix D we get $h=0$. 
Further, we find
\eqn\rxfive{q_{,r}={C\over r(1+r^2)}} (Note that in the notation of 
Appendix D we
have $Q=2$ for the equation that we are solving.) For a solution 
regular at $r=0$
we take $q_{,r}=0$, and thus set $ q=1$. Then we get $f=(1+2r^2)$. 
Note that near
$r=0$ we have
$A_\phi=A_u-A_v=f-q=O(r^2)$, so that the connection is regular at $r=0$. Near
$r=\infty$,  $A_u$ dominates over $A_v$, and we thus get a connection at the
boundary that is a  left moving 1-form on the boundary.

If we had taken instead the condition  $(*d+4)A=0$ from \rsixt\ then 
we would find
that the solution that is large at infinity is not chiral. For this 
reason it appears
reasonable to identify the solution of $(*d-2)A=0$ as the function dual to the
constant mode of the current operator in the boundary.

Having solved for the field $A_\mu$ we can recover the field $B_\mu$ from the
relation \rtw . Thus we observe that we can obtain solutions for the 
fields $A_\mu,
B_\mu$ that go over to a conatant value of the connection at the boundary, and
give solutions that are smooth inside the $AdS_3$ spacetime. These 
fields lift to a
smooth geometry of $AdS_3\times S^3$ when we recall the relation between the
gauge group index and the vector fields on $S^3$, listed in Appendix A. We have
also found along the way excitations that are pure gauge, in eqn 
\rtw ; we take
$\Lambda\ne 0$ but $F=G=0$. We will look at such solutions (at the
nonperturbative level) in the next section, but note here that such 
locally pure
gauge connections are typically singular at $r=0$, in contrast to the smooth
solutions listed in this subsection.

\subsec{Obtaining the field equations from 10-D supergravity}

The equations \rfour \rfive\ were obtained in \deger\  for a general $N=4b$
supergravity theory in D=6. Let us see how we get such a set of equations from
dimensional reduction of Type IIB supergravity in $D=10$, compactified on
$AdS_3\times S^3\times M_4$. The relevant part of the supergravity action is
\eqn\mome{S=\int d^{10} x \sqrt{-g} [R-{1\over 2} \partial\phi\partial\phi
-{1\over 3} e^\phi H^2]} where we are using the Einstein metric, $\phi$ is the
dilaton, and $H=dB$. (We have chosen the normalization of $B$ to agree with the
notation of \deger .) The field equation for $\phi$ is
\eqn\mtwo{\triangle\phi={1\over 3} e^\phi H^2}

Thus if $H^2=0$ then we can consistently set $\phi=0$. The background \eone -
\ethree\ satisfies
\eqn\mthree{H=*H, ~~~H^2=H\wedge*H=H\wedge H=0} where the $*$ is taken in
$AdS_3\times S^3$.

Let us look for deformations $B\rightarrow B+\delta B$ which continue 
to satisfy
$H=*H$, so that $H^2=0$ and the dilaton can continue to be set to 
zero.  The Einstein
equations give
\eqn\msix{R_{a\mu}=H_{aAB}H_\mu{}^{AB}}

Let the nonzero components of the perturbation be of the form
\eqn\mfour{h_{\mu a}=K_{\mu a}(x)Y^{1,1}_a(y), ~~~B_{a\mu}=Z_{\mu
a}(x)Y^{1,1}_a(y)}
\eqn\mfive{\epsilon_a{}^{bc}\partial_bh_{\mu c}=2h_{\mu a},
~~~~\epsilon_a{}^{bc}\partial_bB_{c\mu }=2h_{a\mu }}

Then we get to lowest order in the perturbation
\eqn\mseven{\eqalign{R_{a\mu}&=-{1\over 2}[ h_{\mu a;A}{}^{;A}+h_A^A{}_{;\mu
a}-h_{\mu A;a}{}^{;A}-h_{aA;\mu}{}^{;A}]\cr &=-{1\over
2}[K_{\mu;\nu}{}^{\nu}-K_\nu{}^{;\nu}{}_\mu-2K_\mu]\cr }}

On the other hand
\eqn\meight{H_{\mu AB}H_a{}^{AB}= \epsilon_{\mu}{}^{\nu\lambda}[B_{a\nu,
\lambda} - B_{a\lambda, \nu}]+[B_{\mu b,c}-B_{\mu c,b}]\epsilon_{a}{}^{bc}=
2\epsilon_{\mu}{}^{\nu\lambda}Z_{\nu, \lambda} -4 Z_{\mu }} where we have
expanded to lowest order in the perturbation,  and used the 
background values of
$H$   and  \mfive . Thus we obtain from \msix\ the equation
\eqn\mnine{\triangle K_\mu - K_\nu {}^{;\nu}{}_\mu-2
K_\mu+4\epsilon_{\mu}{}^{\nu\lambda}Z_{\nu, \lambda} -8 Z_{\mu }=0}

Now let us look at the condition $H=*H$. The relevant component of 
this equation is
\eqn\mten{H_{\nu\lambda c}=(*H)_{\nu\lambda c}}

But
\eqn\mel{H_{\nu\lambda c}=\p_\nu B_{\lambda c}-\p_\lambda B_{\nu c}}
\eqn\mtw{(*H)_{\nu\lambda c}= {1\over 6}\epsilon_{\nu\lambda c ABC}H_{A'B'C'}
g^{AA'}g^{BB'}g^{CC'}=\epsilon_{\nu\lambda \mu}h^{\mu}_
c+2\epsilon_{\nu\lambda }{}^\mu B_{c\mu}}

Contracting  \mten\ with $\epsilon_\mu{}^{\nu\lambda}$ (and using \mfive )  we
get
\eqn\mthir{\epsilon_\mu{}^{\nu\lambda}\p_\nu Z_\mu+2Z_{\lambda}+K_\mu=0}

Thus we get the equations in \deger .\foot{In \bal\ it was argued 
that to get a consistent
  dimensional
reduction from 6-D to 3-D we must use both self-dual and anti-self 
dual parts of the
  2-form gauge field.
This does not contradict the fact that we can obtain a class of 
solutions using only the
self-dual
part of the field and obtain for these
the equations in \deger ;  we are investigating these special 
solutions to obtain the dual of
  a constant
  background potential in the CFT.}
  If we had used the  perturbations $K^{1, -1},
Z^{1,-1}$ instead then we would get the lower signs in \rfour , \rfive .

To evaluate correlation functions using the AdS/CFT correspondence we must
evalaute the value of the action of the solution of the supergravity 
equations. The
action to quadratic order is
\eqn\mfourt{S=\int d^3 x \sqrt{-g}[-{1\over 4} F_{\mu\nu} F^{\mu\nu}-2A_\mu
A^\mu +4(-2B_\mu B^\mu -{1\over 4} G_{\mu\nu}G^{\mu\nu}\pm 2 A_\mu
B^\mu -{1\over 2}
\epsilon^{\mu\nu\lambda} A_\mu G_{\nu\lambda})]} where $F=dA, G=dB$. The
first two terms come from the dimensional reduction of $\sqrt{-g}R$ in 6-D, and
the others come from reduction of $-{1\over 3} H^2$.  The upper and lower signs
correspond to the modes $Y^{1, \pm 1}$ in the harmonic expansions on $S^3$. We
are assuming here that we excite fields in a way such that
$H$ remains self-dual,  so that the dilaton is not excited.

The above action should also yield the anomalies of the CFT, 
following the method
used for the case $AdS_5\times S^5$ in \wittenone .

\newsec{`Locally pure gauge'  deformations}

\subsec{Different types of deformations}

In our analysis of section 2.1 we saw that there were locally pure 
gauge excitations
given in \rfourt\ where the field strengths satisfied $G=F=0$. These 
solutions can
be nevertheless nontrivial if we have a nonzero value for the integral of the
connection around a cycle. There are no nonvanishing cycles in 
$AdS_3$, but if we
allow
$r=0$ to be a singular point, then we can find nontrivial flat 
connections. Such flat
conections can be locally gauged away, so locally the spacetime must look like
$AdS_3\times S^3$ though globally it need not reduce to this product form.

Thus we wish to look at the possible ways that one can have a space that looks
locally the same as
$AdS_3\times S^3$; with this analysis we will then not be constrained 
to the linear
order of `locally pure gauge' deformations.  For simplicity of 
presentation let us
first look
  at a `toy model' where the space is locally $AdS_3\times S^1$. Let 
us therefore
start with a `product metric'
\eqn\aone{ds^2=-(1+r^2) dt^2+ {dr^2\over1+r^2}+ r^2 d\phi^2 + d\chi^2} where
$0\le \chi <2\pi$ is a circle of unit length.  To get a space that 
looks locally the same
we can do three things:
\bigskip (a) \quad {\it `Twists'}\quad We can write the metric as
\eqn\aonep{ds^2=-(1+r^2) dt^2+ {dr^2\over1+r^2}+ r^2 d\phi^2 + (d\chi-\alpha
d\phi)^2} with $\alpha$ a constant. We then say that the variables 
$\phi$, $\chi$
are still the ones that are well defined in the sense that
$\phi\rightarrow\phi+2\pi$ and $\chi\rightarrow \chi+2\pi$ give the
identifications of the spacetime. (Thus even though with the definition
$\chi'=\chi-\alpha\phi$ the metric \aonep\ looks locally like \aone ,  the two
metrics are really different.) In a `Kaluza-Klein' ansatz where the coordinate
$\chi$ denotes an internal direction, we would write the metric as
\eqn\atwo{\hat g_{ A B}=\pmatrix{g_{\chi\chi}&A^\chi_\mu g_{\chi\chi}\cr
A^\chi_\mu g_{\chi\chi}& g_{\mu\nu}+A^\chi_\mu A^\chi_\nu g_{\chi\chi}\cr}}
(see for example \ref\mahs{J.~Maharana and J.~H.~Schwarz, ``Noncompact
symmetries in string theory,'' Nucl.\ Phys.\ {\bf B390}, 3 (1993)
[hep-th/9207016].}). Here $A,B$ are coordinates of the 4-d space in 
\atwo , $\hat
g _{AB}$ is the metric \aonep\ and $\mu,
\nu$ are the coordinates
$t,r,\phi$. The 1-form gauge connection is $A^\chi_\mu$, with
$F_{\mu\nu}=\partial _\mu A_\nu^\chi - \partial_\nu A^\chi_\mu$. We find
$A^\chi_\phi=\alpha$,  so the potential is locally pure gauge. The 
effective 3-d
metric is then
\eqn\athree{g_{\mu\nu} dx^\mu dx^\nu=-(1+r^2) dt^2+ {dr^2\over1+r^2}+ r^2
d\phi^2 } Note that $g_{\chi\chi}=1$, so the scalar corresponding to 
the  length of
the compact circle is not excited. The 4d metric \aonep\ is not in 
general regular at
$r=0$. (It is regular if $\alpha$ is an integer $n$ in which case the 
deformation can
be undone by a coordinate transformation.)

We will call this type of deformations `twists'.
\bigskip (b)\quad {\it `Magnetic solutions'}\quad We can write a metric
\eqn\aonepp{ds^2=-(1+r^2) dt^2+ {dr^2\over1+r^2}+ r^2 (d\phi+\beta d\chi) ^2 +
d\chi^2} where again the coordinates $\phi, \chi$ are periodic with 
period $2\pi$.
Now  $g_{\chi\chi}=1+r^2\beta^2$, so the scalar corresponding to the 
size of the
compact circle is excited, and in fact grows monotonically with 
increasing $r$. The
gauge potential is $A^\chi_\phi=(g_{\chi\chi})^{-1}g_{\chi\phi}=\beta {r^2\over
1+r^2\beta^2}$.  There is a nonvanishing magnetic field strength 
$F_{r\phi}$. The
effective metric $g_{\mu\nu}$ in the Kaluza-Klein ansatz is
\eqn\athreep{g_{\mu\nu} dx^\mu dx^\nu=-(1+r^2) dt^2+ {dr^2\over1+r^2}+
{r^2\over 1+r^2\beta^2} d\phi^2 } (This is the `string metric' which 
has a gravity
action $\int \sqrt{-g}e^{-2\phi}R+\dots$ where $\phi=-{1\over 4} \log
g_{\chi\chi}$.) Note that at $r\approx 0$ the metric \aonepp\ 
approaches \aone ,
and so is regular at $r=0$.  If we set the $t,r$ part of the metric to equal
$-dt^2+dr^2$ then this metric would be equivalent to the one appearing in the
Melvin solution\foot{I thank A. Tseytlin for pointing out these references.}
\ref\melvin{M.A. Melvin, {\it Phys. Lett.} {\bf 8} (1964) 65, J.~G.~Russo and
A.~A.~Tseytlin, ``Magnetic flux tube models in superstring theory,'' 
Nucl.\ Phys.\
{\bf B461}, 131 (1996) [hep-th/9508068].}.
\bigskip

(c)\quad We can take the metric \aone\ but include a `conical 
defect'; i.e.,  let the
range of $\phi$ be $0\le \phi < 2\pi\gamma$.  We can define a new coordinate
$\phi'=\phi/\gamma$ that still runs over $0\le \phi' < 2\pi$ by writing
\eqn\afour{\eqalign{ds^2&=-(1+r^2) dt^2+ {dr^2\over1+r^2}+ r^2\gamma^2
d\phi'^2 + d\chi^2\cr &=-(\gamma^2+(r\gamma)^2) d(t/ \gamma)^2+
{d(r\gamma)^2\over \gamma^2+(r\gamma)^2}+ (r\gamma)^2 d\phi'^2 +
d\chi^2\cr
  &=-(\gamma^2+r'^2) dt'^2+ {dr'^2\over \gamma^2+r'^2}+ r'^2 d\phi'^2 +
d\chi^2\cr }} where  we have defined $r'=r\gamma, t'=t/\gamma$ in the last
step.\foot{Writing $\gamma^2=\delta$, we can continue \afour\ to $\delta=-A$.
Defining
$\tilde
\phi=it,
\tilde t=i\phi$, $\tilde r^2=r^2-A$,  it is interesting that we get
$$ds^2=-(r^2-A) dt^2+ {dr^2\over r^2-A}+ r^2 d\phi^2 =-(\tilde r^2+A) d\tilde
t^2+ {d\tilde r^2\over \tilde r^2+A}+ \tilde r^2 d\tilde \phi^2 $$ 
Further,  after
allowing rotations in $t, \phi$ it would be interesting to 
investigate the possible
relation of such metrics to the `kinks' observed in thermal strings in
\ref\mathur{S.~D.~Mathur, ``Real time propagator in the first quantized
formalism,'' Proceedings of the third Thermal Fields Workshop held at  Banff.
hep-th/9311025.}.}

It can be seen that we can `mix' the deformations of types (a), (b), 
(c) to make
other spacetimes. Thus we can take an arbitrary linear combinations of
$\phi,
\chi$ (with constant coefficients) to multiply the terms in \aone\ 
with coefficients
$r^2$ and unity, and for any such construction we can limit the range of the
coordinate $\phi$  as in (c).

\subsec{`Twists' for $AdS_3\times S^3$}

The metric of the undeformed $AdS_3\times S^3$ is
\eqn\qtenp{ds^2= -\cosh^2\rho dt^2+d\rho^2+\sinh^2\rho d\phi^2+
d\Omega_3^2} where we have taken unit radius for the $AdS$ and sphere factors.
The vector field $A_\mu$ in the above section arose from $h_\mu^a$, and a
corresponding perturbation of the 2-form field $B$. A nonzero value for
$h_\mu^a$ indicates that the $AdS$ space is not orthogonal to the 
sphere, though
since we are trying to make a flat connection, locally it should be 
possible to choose
coordinates where no effect of the gauge field is seen at all.

Let the sphere be given by
\eqn\tfour{x_1^2+x_2^2+x_3^2+x_4^2=1} Let
\eqn\tfive{\eqalign{x_1&=q\cos\chi\cr x_3&=q\sin\chi\cr
x_2&=\sqrt{1-q^2}\cos\psi\cr x_4&=-\sqrt{1-q^2}\sin\psi\cr}}

In Appendix A we list the vector fields on $S^3$ that correspond to the
decomposition $so(4)=su(2)\oplus su(2)$. Let $M_{ij}$ be the  generators of the
rotations of the Cartesian coordinates $x_i$ of the sphere (see 
Appendix A ). Let us
make the analogue of the deformation of type (a) in the above 
subsection. We wish
to choose a fixed generator of rotations of this
$S^3$, and rotate the $S^3$ at a constant rate as we increase the 
$AdS$ coordinate
$\phi$. We can take any linear combination of the generators $M_{ij}$, or
equivalently, of the left and right $su(2)$ generators $J^a, K^a 
~~(a=1,2,3)$. For
presentational purposes the simplest case is where we have an equal 
magnitude of
rotation in the two $su(2)$ factors. We can choose coordinates on the sphere so
that these rotations are in the directions $J^3, K^3$, and then for 
equal rotations
in the two $su(2)$s  the generator of rotations is propotional to
\eqn\ttwo{J_3+K_3={1\over 2} [M_{13}+M_{42}]+{1\over
2}[M_{13}-M_{42}]=M_{13}\rightarrow {\partial\over \partial\chi}}

Thus we get a rotation of the circle $\chi$ as we increase the $AdS$ coordinate
$\phi$, so this is similar to the situation in \aonep . But note that 
the radius of the
$\chi$ circle $q$ is not a constant but varies in the range $0\le q\le 1$.

We can find coordinates $\phi', \chi'$ in the $\phi, \chi$ space 
where the metric in
this subspace is
\eqn\tseven{g'=\pmatrix{g'_{\chi'\chi'}&0\cr 0&g'_{\phi'\phi'}\cr
}=\pmatrix{q^2&0\cr 0&r^2\cr } } (The other coordinates are orthogonal to these
two and their metrics are left unchanged.)  The coordinate $\phi'$ 
still labels the
different spheres over different points on a fixed $\rho$, fixed $t$ 
circle in the
$AdS$ space, but the coordinate $\chi'$ is not a periodic one in the 
spacetime. Let
the periodic coordinate be instead
\eqn\teight{\chi=\chi'-\alpha\phi', ~~~\alpha={\rm constant}} Writing
\eqn\tnine{\phi=\phi'} we get the metric in the $\phi, \chi$ subspace in the
unprimed coordinates
\eqn\tten{g\equiv \pmatrix{g_{\chi\chi}& g_{\chi\phi}\cr  g_{\phi\chi}&
g_{\phi\phi}\cr} =\pmatrix{g'_{\chi'\chi'}&\alpha g'_{\chi'\chi'}\cr \alpha
g'_{\chi'\chi'}& g'_{\phi'\phi'}+\alpha^2g'_{\chi'\chi'}\cr}} The 
determinant is
unchanged
\eqn\ttw{det g = det g' = g'_{\phi'\phi'}g'_{\chi'\chi'}} The inverse 
metric in this
subspace is
\eqn\tel{(g)^{-1}=\pmatrix{{1\over g'_{\chi'\chi'}}+{\alpha^2\over
g'_{\phi'\phi'}}&-{\alpha\over  g'_{\phi'\phi'}}\cr
  -{\alpha \over g'_{\phi'\phi'}}&{1\over g'_{\phi'\phi'}}\cr}}

\subsec{Analogues of the other deformations}

We can also make deformations of $AdS_3\times S^3$ which are analogous to the
deformations of type (b) (`magnetic deformations') and type (c) 
(`conical defects').
the conical defects are made in the same way as for $AdS_3\times 
S^1$, since the
space is a direct product of an $AdS_3$ space with conical defect 
(eq. \afour ) with
the internal space which can now be taken to be $S^3$ instead of $S^1$. To make
the analogue of the `magnetic deformations' we take the circle parametrized by
$\chi$ on
$S^3$ (eq. \tfour ) and use this to replace the  $S^1$ in the 
construction of type (b)
above. Note however that at $q=0$ the size of this circle goes to 
zero, and thus the
spacetime is singular in general.

\subsec{Changing to a chiral gauge}

The above construction gives us a connection
\eqn\jone{A_\phi^{J^3}=A_\phi^{K^3}=\alpha} In the boundary CFT the left movers
can carry the charge  $J^a$ and the right movers can carry the charge $K^a$. We
can change gauge to let the $J^a$ component of $A_\mu$ be of the form
$A_{\phi+t}$ and the $K^a$ component be of the form $A_{\phi-t}$. To do this we
must make a gauge rotation linear in
$t$, such that we get
\eqn\jtwo{A_t^{J_3}=-A_t^{K_3}=\alpha} But
\eqn\ttwop{J_3-K_3={1\over 2} [M_{13}+M_{42}]-{1\over
2}[M_{13}-M_{42}]=M_{42}\rightarrow {\partial\over \partial\psi}} so that the
above gauge transformation corresponds to
\eqn\jthree{\psi=\psi'-\alpha t', ~~~t=t'} analogous to \teight , 
\tnine . The metric
in the $\psi, t$ subspace is
\eqn\tsevenp{g'=\pmatrix{g'_{\psi'\psi'}&0\cr 0&g'_{t't'}\cr
}=\pmatrix{1-q^2&0\cr 0&-(1+r^2)\cr } }
\eqn\ttenp{g\equiv \pmatrix{g_{\psi\psi}& g_{\psi t}\cr  g_{t\psi}& g_{tt}\cr}
=\pmatrix{g'_{\psi'\psi'}&\alpha g'_{\psi'\psi'}\cr \alpha g'_{\psi'\psi'}&
g'_{t't'}+\alpha^2g'_{\psi'\psi'}\cr}}

Note however that the only geometric invariants of these flat 
connections are the
periods of the $J$ and $K$ subgroups around the compact circle $\phi$. The $t$
direction is noncompact, and so we can adjust the generator $\hat 
A_t$ to be any
constant we want by a gauge transformation.

\subsec{Unequal values of the connection in the two subgroups}

We list for completeness the metric that arises from the case where we have a
connection with unequal magnitudes in the left and right $su(2)$ 
factors. We can
choose coordinates to again put the connections along $J^3, K^3$.  Let the
connection in the $\phi$ direction $\hat A_\phi$ have the form
\eqn\jsix{\hat A_\phi=\alpha M_{13}+\beta M_{42}=\alpha(J^3+K^3)+\beta (
J^3-K^3)=(\alpha+\beta)J^3+(\alpha-\beta)K^3} If we wish to make the 
$J$ part of
the connection to be of the form $A_{\phi+t}$ and the $K$ part of the 
connection of
the form $A_{\phi-t}$, then we let
\eqn\jsixp{\hat A_t=(\alpha+\beta)J^3-(\alpha-\beta)K^3=\alpha(J^3-K^3)+\beta (
J^3+K^3)=\alpha M_{42}+\beta M_{13}} Recall that $M_{13}\rightarrow
{\partial\over \partial\chi}, ~~M_{42}\rightarrow {\partial\over 
\partial\psi}$.  We
then get the metric
\eqn\jseven{\eqalign{ds^2=&
-(r^2+1-\beta^2\cos^2\theta-\alpha^2\sin^2\theta)dt^2+(r^2+\alpha^2\cos^2
\theta+\beta^2\sin^2\theta)d\phi^2+{dr^2\over
r^2+1}\cr &+d\theta^2+2(\alpha d\phi+\beta dt)\cos^2\theta d\chi+2(\alpha dt
+\beta d\phi) \sin^2\theta d\psi+\cos^2\theta d\chi^2+\sin^2\theta d\psi^2\cr}}

If we  start instead with a metric having a conical defect (which had 
been put in
the form given by the primed coordinates in \afour ) and perform the twists  as
above, then we would get the more general metric
\eqn\jsevenp{\eqalign{ds^2=&
-(r^2+\gamma^2-\beta^2\cos^2\theta-\alpha^2\sin^2\theta)dt^2+(r^2+\alpha^2
\cos^2\theta+\beta^2\sin^2\theta)d\phi^2+{dr^2\over
r^2+\gamma^2}\cr &+d\theta^2++2(\alpha d\phi+\beta dt)\cos^2\theta
d\chi+2(\alpha dt +\beta d\phi) \sin^2\theta d\psi+\cos^2\theta
d\chi^2+\sin^2\theta d\psi^2\cr}}

\newsec{Localised modes for the `twisted geometries'}

  In \ref\sm{J.~Maldacena and A.~Strominger, ``AdS(3) black holes and a stringy
exclusion principle,'' JHEP {\bf 9812}, 005 (1998) [hep-th/9804085].} it was
shown that if we consider a scalar of mass $m$ in $AdS_3$ then we obtain
normalisable solutions localised near $\rho=0$ (over a scale of the 
order of the
$AdS$ radius) which are dual to primary operators in the boundary CFT  of
dimension $h=\bar h=1/2(1+\sqrt{m^2+1})$.  It is not clear what the 
significance
of the general `twisted geometry' is, so it is not clear how to 
interpret solutions to
the scalar wave equation for general twists. Nevertheless the spectrum of these
modes appears to be interesting and solutions parallel in some ways the
wavefunctions in \sm\  that were dual to chiral primary operators, so we carry
out this computation below.

In our case the 6-d space does not factorize globally into an
$AdS_3$ part and a sphere, so we look at scalars in the 6-d theory. We let the
scalar be massless and minimally coupled at the linear order (there are several
such scalars in the theory). For a given
$R$-charge the lowest dimension CFT operators typically arise from supergravity
fields that have two indices along the $S^3$. Since we are starting 
with  a scalar,
we  are considering supersymmetry descendents of chiral primaries (this is
analogous to the operator $\tr F^2$ dual to the dilaton in $AdS_5\times S^5$).

Consider the metric that corresponds to the connection
$A_\phi^{J^3}=A_\phi^{K^3}=\alpha$:
\eqn\zone{ds^2=-(1+r^2) dt^2+{dr^2\over r^2+1} +d\theta^2+2\alpha
\cos^2\theta d\phi d\chi+(r^2+\alpha^2\cos^2\theta)d\phi^2+\cos^2\theta
\chi^2+\sin^2\theta d\psi^2} The scalar $f$ in 6-d satisfies the equation
\eqn\tone{{1\over \sqrt{-g}}[f_{,M}g^{MN}\sqrt{-g}]_{,N}=0} Consider the ansatz
\eqn\ztwo{f=e^{-i\omega t} e^{i m \phi} Y^{l_1, 0}(y)} where $Y^{l_1,0} $ is a
spherical harmonic for the scalar (the left and right spins must be equal for a
scalar function: $j=k=l_1/2, -j\le j_3\le j, -k\le k_3\le k$), and
$y$ denotes the coordinates on the sphere. With this ansatz we find the wave
equation
\eqn\fonep{{1\over r} [ f_{,r} r (1+r^2)]_{,r}+ \omega^2 (1+r^2)^{-1}f-
[l_1(l_1+2)+(\alpha(j_3+k_3)-m)^2r^{-2}]f=0} In Appendix E we bring 
this equation
to a hypergeometric form that parallels the equation for the 1-form 
fields studied
in section 2.

In \sm\ the wavefunctions dual to the chiral primaries had the form
\eqn\ttwop{f=e^{-i\omega t}(\cosh\rho)^{-\mu} Y^{l_1, 0}(y)} where $r=\sinh
\rho$.  To get the generalisation of these special solutions  to the twisted
geometries consider the ansatz
\eqn\ttwo{f=e^{-i\omega t}(\cosh\rho)^{-\mu} 
(\sinh\rho)^{-\nu}Y^{l_1,0}(y)} (We
have added a parameter $\nu$ since as we will see we will encounter a term
singular at $\rho=0$ in our case.)

Substituting in the wave equation we find that  we need to have vanish the
expression
\eqn\teight{\eqalign{\omega^2 & (\cosh\rho)^{-2-\mu}(\sinh\rho)^{-\nu}+\mu^2
(\cosh\rho)^{-\mu-2}(\sinh\rho)^{-\nu+2}\cr &+(\mu(\nu-2)+\nu(\mu-2))
(\cosh\rho)^{-\mu}(\sinh\rho)^{-\nu}+\nu^2 (\cosh\rho)^{-\mu+2}
(\sinh\rho)^{-\nu-2}\cr & -[2j(j+1)+2 k(k+1)] (\cosh\rho)^{-\mu}
(\sinh\rho)^{-\nu}-\alpha^2(j_3+k_3)^2 (\cosh\rho)^{-\mu}
(\sinh\rho)^{-\nu-2}\cr }}

The most singular terms at the origin must cancel, so we set
\eqn\tten{\nu^2=\alpha^2(j_3+k_3)^2} Let us take $\alpha>0$. We get two choices
\eqn\tel{\nu=\pm \alpha (j_3+k_3)} Equating the coefficients of the other
independent expressions to zero we get
\eqn\tthir{-\mu^2+\omega^2=0, ~~~\omega=\pm \mu} (We will take $\omega>0$).
and
\eqn\tfourt{\mu^2+\nu^2+\mu(\nu-2)+\nu(\mu-2)-l_1(l_1+2)=0  }

Then \tfourt\ gives
\eqn\tfift{(\mu+\nu)(\mu+\nu-2)-l_1(l_1+2)=0} Thus we get two choices
\eqn\tsixt{\mu+\nu=l_1+2, ~~~\mu+\nu = -l_1} For a convergent solution at
infinity we need $\mu+\nu>0$. So we discard the second choice. Then we have
\eqn\tsevent{\omega= \pm\mu = \pm[(l_1+2)\mp \alpha(j_3+k_3)]}

  Note that states with negative $\nu$ are singular at $\rho=0$ while those with
positive $\nu$ are regular. It is interesting that we get solutions 
that are regular
both at $r=0$ and at $r=\infty$.

If we go to a gauge where the connections for the two gauge groups 
are purely left
moving or   purely right moving forms then we get the metric \jseven\ (with
$\beta=0$). The solutions to the wave equation are of course the same as above
since we have just made a coordinate transformation. However in this new frame
the frequencies can be read off from the dependence
\eqn\zsix{e^{-i\omega t}e^{i (j_3-k_3)\psi}\equiv e^{-i\omega t'}e^{i
(j_3-k_3)(\psi'+\alpha t')}=e^{-i(\omega-\alpha(j_3-k_3)) t'}e^{i 
(j_3-k_3)\psi'}} so
that
\eqn\zseven{\omega'=\omega-\alpha(j_3-k_3)} and in particular for the modes
\tsevent\
\eqn\zeight{\omega'= \pm(l_1+2)-2 \alpha j_3, ~~~\omega'=\pm (l_1+2)+2 \alpha
k_3} for the two signs respectively in \tsevent .

\subsec{Including a `conical defect'}

Let us take the metric \zone\ but let the range of $\phi$ be $0\le 
\phi \le 2\pi
\gamma$. In the ansatz \ztwo\ we need just take $m=m'/\gamma$ where
$m'$ is an integer, and the rest of the computation for the frequencies is
unaffected. But it is also interesting to see the effect of the 
conical defect on the
specific modes \ttwo . There is no explicit $\phi$ dependence, so the 
frequencies
\tsevent\ in the coordinate $t$ are unaffected.  But $\phi, t$ are 
not asymptotically
AdS coordinates now, and  to change to asymptotically AdS coordinates we must
write $\phi'=\phi/\gamma, t'=t/\gamma$ (see eq. \afour ). Thus in these
coordinates $\phi', t'$ the frequencies are given through
\eqn\zten{e^{-i\omega t}=e^{-i\omega't'}=e^{-i\omega' t/\gamma},
~~~\omega'=\omega\gamma} Thus the frequencies in the asymptotically AdS
coordinates are not in general integral multiples of the natural AdS frequency.

\newsec{Metrics from the rotating string}

In \bal , \mm\ a set of metrics was constructed that corresponded to 
the geometry
of a D1-D5 system carrying angular momentum. The string in 6-D (after
compactifing spacetime on a $T^4$) is wrapped on the circle $\phi$ and the
fermions have periodic boundary conditions on this circle. Thus the 
D1-D5 system
has Ramond type boundary conditions.  The metric as given in \mm\ is (after
adapting the notation to the one we use)
\eqn\lonep{\eqalign{{ds^2\over \sqrt{k}}&={1\over h}[-dt^2+d\phi^2]
+hf[d\theta^2+{r^2 dr^2\over (r^2+\gamma_1^2)(r^2+\gamma_2^2)}]\cr
&+{2\over hf}[(\gamma_2 dt +\gamma_1 d\phi) \cos^2\theta d\chi+(\gamma_1 dt
+\gamma_2 d\phi)\sin^2\theta d\psi]\cr
&+[h(r^2+\gamma_2^2)+(\gamma_1^2-\gamma_2^2){\cos^2\theta\over
hf^2}]\cos^2\theta
d\chi^2+[h(r^2+\gamma_1^2)-(\gamma_1^2-\gamma_2^2){\sin^2\theta\over
hf^2}]\sin^2\theta d\psi^2, \cr & f=r^2+\gamma_1^2\cos^2\theta +\gamma_2^2
\sin^2\theta ,\cr &h={\sqrt{k}\over R_y^2}(1+{R_y^2Q_1\over
kf})^{1/2}(1+{R_y^2Q_5\over kf})^{1/2}\cr }} Here $k=N_1N_5$, where $N_1, N_5$
are the numbers of D1 and D5 branes.

For equal rotations in the left and right $su(2)$ factors the near 
horizon geometry
is
\eqn\lone{\eqalign{{ds^2\over
\sqrt{k}}&=-(r^2+\gamma_1^2\cos^2\theta)[dt^2-d\phi^2] +{dr^2\over
r^2+\gamma_1^2}\cr &+2\gamma_1d\phi d\chi\cos^2\theta+2\gamma_1 dt
d\psi\sin^2\theta+\cos^2\theta d\chi^2+\sin^2\theta d\psi^2\cr}}

The coefficient of $dt^2$ is
\eqn\ltwo{-(r^2+\gamma_1^2\cos^2\theta)=
-(r^2+\gamma_1^2-\gamma_1^2\sin^2\theta)}
Thus the metric \lone\ is of the form \jsevenp\ if we have a conical 
defect with
$\gamma=\gamma_1$ and perform `twists' with  $\alpha=\gamma_1$, $\beta=0$.
   (The case $\gamma_1\ne \gamma_2$ of \mm\ is not of the form \jseven\ since
the metric differs by a shift in the coefficient of $d\phi^2$  of the form
$r^2\rightarrow r^2+C$.)

\subsec{Microscopics of the D1-D5 system}

To comment on some interesting aspects of the solution in \bal , \mm\ we recall
the microscopics of the D1-D5 system. We can think of the D1-D5 bound 
state as a
set of $N_1$ D1-branes living inside $N_5$ D5-branes. Each D-string in the
5-branes can be regarded as being `fractionated' into $N_5$ substrings, giving
$N_1N_5$ substrings in all. Each  substring can be placed in a state 
which is spin
$(0,0)$ under $su(2)\oplus su(2)$, in a state which is
$(1/2, 1/2)$, or into fermionic states that are $(1/2,0)+ (0, 1/2)$. 
Two or more
substrings can be joined into a `long string', but this long string 
can carry only the
spins that a single substring could have carried. Thus in particular 
if we join all the
substrings into one long string then we can only get a spin of order 
unity, which
will not be visible at the classical level. On the other hand if we 
do not join any of
the fractionated strings to each other, and orient the spins of each 
to all lie in the
same direction, then we can get a state with left and right angular 
momenta $(j_L,
j_R)=(k/2, k/2)$, where $k=N_1N_5$.

(One way to see the above properties is to perform a sequence of S 
and T dualities
(along the compact $T^4$ and $\phi$  directions) such that the $N_5$ D5-branes
and
$N_1$ D-strings transform to an $N_5$ times wound elementary string carrying
$N_1$ units of momentum. The $N_1$ units of momentum can be carried by
oscillators (acting on the elementary string winding state) which each carry a
momentum in multiples of $1/N_5$\ref\dmfrac{S.~R.~Das and S.~D.~Mathur,
``Excitations of D-strings, Entropy and Duality,'' Phys.\ Lett.\ {\bf 
B375}, 103
(1996) [hep-th/9601152].}, and  a  vector index of the space transverse to the
string; this index  can be chosen to give spin 1 in the 
$so(4)=su(2)\oplus su(2)$
which must be used to describe the angular momenta of the system.)

(a)\quad First we note that the D1-D5 system that has been given an angular
momentum $(j_L, j_R)=(k/2, k/2)$ (with no momentum charge) is in a different
`phase' from the D1-D5 system which gives the D1-D5-momentum black holes. In
the 3-charge black hole, the microscopic state that is most favoured 
is one where
the substrings of the D1-D5 system all join up into one long string
\ref\malsuss{J.~M.~Maldacena and L.~Susskind, ``D-branes and Fat Black Holes,''
Nucl.\ Phys.\ {\bf B475}, 679 (1996) [hep-th/9604042].} (thus reducing their
own angular momentum to essentially zero) and then any angular momentum
that the system may have is carried by momentum modes that run up and down
the string (the left moving momentum modes carry $j_L$, the right moving ones
carry $j_R$). The 3-charge extremal hole has only left moving momentum modes,
and so we only have $j_L$. In the limit that we reduce this momentum 
to zero, we
get no momentum modes to carry any angular momentum, and so we are forced
to
$j_L=0, j_R=0$.  This fact is mirrored in the fact that the left and right
temperatures of the hole (which combine to  give the Hawking temperature) are
complex for the classical configuration that has only D1 and D5 charges, unless
$j_L=j_R=0$. (The general expressions for these temperatures can be found in
\ref\cl{M.~Cvetic and F.~Larsen, ``General rotating black holes in 
string theory:
Greybody factors and  event horizons,'' Phys.\ Rev.\ D {\bf 56}, 4994 (1997)
[hep-th/9705192].}.)

But from the microscopic picture we can see that starting from the 3-charge
system we can reduce the momentum (and the angular momentum) to zero, but
then break up the joined substrings into separate substrings  (which 
can now each
carry a spin $(1/2, 1/2)$ under $su(2)\oplus su(2)$); thus the system acquires
angular momentum by a different mechanism from the three charge hole.  Closely
related to this phenomenon are the   `phase transitions' where the 
substrings join
or split to maximise entropy \ref\mathu{S.~D.~Mathur, ``Emission rates, the
correspondence principle and the information  paradox,'' Nucl.\ 
Phys.\ {\bf B529},
295 (1998) [hep-th/9706151].} (see also \ref\cvetic{M.~Cvetic and S.~S.~Gubser,
``Thermodynamic stability and phases of general spinning branes,'' JHEP{\bf
9907}, 010 (1999) [hep-th/9903132], J. Maldacena, {\it talk at Strings 99,
Potsdam}.}).

(b) \quad Now we look at some interesting energy and angular momentum scales
that appear from the geometry of \mm , and find corresponding scales in the
microscopic D1-D5 theory. This analysis is rough, and it is not 
possible to claim
from this that there is in fact an identification between the 
physical quantities
being compared. But it is nevertheless interesting to see related 
scales appearing
in the two pictures. We will look at the case where the rotation parameters of
\mm\ are $\gamma_1=1,
\gamma_2=0$, which gives a nonsingular spacetime.

If we take $R_y$  large then we find that we have a large region that is
approximately $AdS_3\times S^3$, and can relate the analysis around 
$r=0$ in the
full metric of the spinning string to quantities computed in the 
$AdS$ geometry. In
what follows we assume that such a limit has been taken.

The near horizon geometry of the D1-D5 system is governed by the number
$N_5$. Thus a giant graviton made of a D-string has an angular momentum
$N_5$ units \ref\gg{J.~McGreevy, L.~Susskind and N.~Toumbas, ``Invasion of the
giant gravitons from anti-de Sitter space,'' JHEP{\bf 0006}, 008 (2000)
[hep-th/0003075], M.~T.~Grisaru, R.~C.~Myers and O.~Tafjord, ``SUSY 
and Goliath,''
JHEP{\bf 0008}, 040 (2000), A.~Hashimoto, S.~Hirano and N.~Itzhaki, ``Large
branes in AdS and their field theory dual,'' JHEP{\bf 0008}, 051 (2000)
[hep-th/0008016],  [hep-th/0008015]. S.~R.~Das, A.~Jevicki and S.~D.~Mathur,
``Giant gravitons, BPS bounds and noncommutativity,'' Phys.\ Rev.\ D {\bf 63},
044001 (2001) [hep-th/0008088],  S.~R.~Das, A.~Jevicki and S.~D.~Mathur,
``Vibration modes of giant gravitons,'' Phys.\ Rev.\ D {\bf 63}, 024013 (2001)
[hep-th/0009019] }.  This is the same as the maximum spin that can be achived
by a D1 brane when it dissolves in the D5 branes. There is a `twist' 
between the
coordinates natural to the AdS region and natural at infinity. Thus while the
angular momentum of the giant graviton in the AdS region can have any
orientation, these differently oriented states have different angular 
momenta as
seen from infinity, depending on the relation between the direction 
of rotation and
the direction of the `twist' in the geomery.  In particular consider the
configuration where the D string is wound on the circle $\phi$, and 
is extracted
away from the AdS region to infinity. Here we can have a static 
configuration that
will have no angular momentum. This appears to be a reflection of the fact that
when we extract a D1 brane from the D5 branes then it can no longer have a spin
greater than unity.

(c) \quad Note that between $r=0$ and $r=\infty$ there is a redshift
\eqn\lseven{{g_{tt}^{1/2}(r=\infty)\over g_{tt}^{1/2}(r=0)}={h^{1/2}(r=0)\over
h^{1/2}(r=\infty)}\approx {R_y\over k^{1/4}}} where at $r\approx 0$ we have
written $r^2+\cos^2\theta\approx 1$.  In section 4 we had seen that there are
localised solutions to the scalar wave equation in the `twisted' AdS 
spaces. If the
full geometry has a large $AdS$ type region then there can be long lived modes
localised near $r=0$.  Such a mode has a wavelength of order the $AdS$ radius
$\sim k^{1/4}$, and thus a frequency in the $t$ coordinate 
$\omega\sim k^{-1/4}$.
After taking into account the redshift, this gives an energy as seen 
from infinity
\eqn\leight{E\sim  k^{-1/4}({R_y\over k^{1/4}})^{-1}\sim {1\over R_y}} But now
note that from the microscopic view of the D1-D5 system at weak coupling, the
energy of excitation of the substrings is also $\sim 1/R_y$; since in 
the state with
$(j_L, j_R)=(k/2, k/2)$ all the substrings are separately wound on a 
circle of length
$R_y$ and not joined together into longer strings.

The scalar wavefunctions localised in the $AdS$ region have frequencies in the
coordinate $t$ that satisfy $\omega\ge 1$,  so that they become 
travelling waves at
$r=\infty$. Thus any wavepacket localised near $r=0$ will escape to 
infinity. But as
we increase $R_y$, we will have a wavefunction that decays for a 
longer distance
in the AdS part of the geometry, thus having a smaller overlap with 
the travelling
wave at infinity, and thus a longer lifetime. This agrees 
qualitatively with the
microscopic intuition that excitations on a longer string take longer 
to `collide' and
escape as radiation.

(d)\quad Now consider a D-string that is dissolved in the $N_5$ D5 
branes, and let
its $N_5$ substrings join up into one long string.  Then we expect 
the microscopic
picture to yield an energy of excitation
\eqn\lten{E\sim {1\over R_y N_5}} Let us  see if we can find such an 
energy scale
anywhere in the classical geometry.  The D-string under consideration is not
joined to other substrings in the system, and can be thus a giant graviton
wrapped in the $AdS$ region. The mass of this giant graviton is $N_5$ times the
natural energy scale
$R_{AdS}^{-1}$ of the $AdS$ space. Since $R_{AdS}\sim k^{1/4}$, we get for this
mass $M=N_5 k^{-1/4}$. For $AdS_3$ the giant graviton does not have a radius
that is fixed by its angular momentum; rather the angular momentum is fixed and
there is no potential for the variable that gives the radius of the 
D-string. Thus we
can have a slow variation of this radius, getting the physics of a 
massive particle
that is moving without potential. Let us set the momentum of this 
radial motion to
be $P\sim 1/R_{AdS}\sim k^{-1/4}$. Then the energy of the motion is $P^2/(2M)$.
Taking into account the redshift
\lseven\ we get the energy as seen from infinity
\eqn\lel{E\sim ({R_y\over k^{1/4}})^{-1}{P^2\over M}\sim ({R_y\over
k^{1/4}})^{-1}{ k^{-1/2}\over N_5 k^{-1/4}}\sim {1\over R_y N_5}} which agrees
with \lten .

\newsec{Discussion}

We have examined the equations of the fields in supergravity that have the
correct quantum numbers to couple to the R-currents of the boundary CFT, and
argued that these fields must be used to get the supergravity description
corresponding to  adding a constant  external gauge potential to the 
boundary CFT.
Such an external gauge field effectively changes the boundary condition on the
fermions, and thus can be used to generate a 1-parameter family of 
CFTs starting
with the CFT in the NS sector, which is dual to $AdS_3\times S^3$.  We have not
followed the smooth solutions beyond the lowest order in the fields, 
but based on a
comparison of quantum numbers we expect that for boundary values of the
supergravity fields that represent flat connections in the CFT we can 
find solutions
at the nonlinear level using only the fields that we have considered here.

But in this process we argued that one must look at smooth solutions of these
supergravity fields in the interior of $AdS$ space. (As we follow the family
  of configurations to larger values of the connection, at some point 
the solution may
of course become singular as a solution of supergravity.) We can on the other
hand look at `Wilson lines in the bulk' which are deformations of the 
$AdS_3\times
S^3$ such that locally the space is still $AdS_3\times S^3$. We looked at three
different deformations of this kind, but these are spacetimes which 
are generically
singular.  While singular solutions can have a nonzero value of $A_\phi$ as
$r\rightarrow 0$ (which can at special values be gauged away to zero) 
the family
of regular solutions has $A_\phi\rightarrow 0$ as $r\rightarrow 0$.

The proposal of \wadia\ was to represent spectral flow in the CFT (which
corresponds to changing boundary conditions of the fermions) by 
putting a Wilson
line of the gauge field in the bulk which couples to the R-current, while also
including a conical defect.\foot{ In \town ,\bal\  a similar 
construction of Wilson lines and
conical defects was used, but in these references the goal was to construct
supersymmetric conical defects. In \bal\  the conical defect and 
`Wilson lines in the
bulk' were argued to be related to  spectral flow, though in a manner somewhat
different from \wadia ; the singular spacetimes  were argued to be ensembles
  of states in the R sector.}
The singular spacetimes
could represent the supergravity solutions required to spectral flow 
the theory if the action
on these configurations was lower than the action on the smooth 
configurations; checking
this however would require resolving the singularity in string theory and then
computing the resulting action. More generally, we have to 
distinguish states of the CFT
from
deformations of the Hamiltonian of the CFT; the former are localised 
near the center of AdS
while the latter correspond to solutions that grow towards the boundary
\ref\balold{V.~Balasubramanian, P.~Kraus and A.~Lawrence,
``Bulk vs. boundary dynamics in anti-de Sitter spacetime,''
Phys.\ Rev.\ D {\bf 59}, 046003 (1999)
[hep-th/9805171].}. The addition of a
background gauge
potential to the CFT is a deformation of the Hamiltonian. The
issue of singular versus regular configurations is certainly one 
deserving further study.

We examined the wavefunctions of a scalar field that were localised near the
center of the $AdS$ spaces with twists and conical defects. If we 
included only a
twist, the shift in the spectrum  \tsevent\ was formally similar to 
the effect of
spectral flow. But after we change to a gauge where the left 
connection  carries a
left moving index and the right connection carries only a right 
moving index, the
form the the frequencies  becomes different.  (It is also not clear 
how to get the
momentum of the state in the
$AdS$ to correspond to the momentum in the CFT.)  If we include conical defects
then the frequencies of these wavefunctions (expressed in a 
coordinates that are
asymptotically $AdS$) are not integer multiples of the natural AdS frequency.

In \mm\ the solution for $\gamma_1=1, \gamma_2=0$ was shown to be
nonsingular. At $r=\infty$   the fermions of the theory are periodic around the
circle $\phi$, which becomes a compact circle of fixed radius.  In 
this region the
gauge connection has gone to zero, so we can say that we have Ramond sector
fermions. As we move to smaller $r$, the fermion wavefunction remains periodic
by continuity, but we develop a gauge connection. When we reach the region that
can be approximated by $AdS_3\times S^3$ then the value of this connection is
$\int  A_\phi d\phi=2\pi$. We can do a coordinate transformation to 
get $A=0$, but
in the process the fermions in the new coordinate system would be antiperiodic
around the $\phi$ circle. This means that they can be smoothly continued to the
origin at
$r=0$. Thus in the region where the spacetime is approximately $AdS_3\times
S^3$  we are in the NS sector in the locally correct coordinates; an 
R fermion at
infinity gives an NS fermion in the AdS region.  Thus this case 
differs from the
spacetime which we would obtain if we try to  deform $AdS_3\times S^3$ to get
the R sector of the CFT.

   In
\ref\lm{O.~Lunin and S.~D.~Mathur, ``Correlation functions for M**N/S(N)
orbifolds,'' hep-th/0006196.} a method was found to compute the correlation
functions of twist operators in the bosonic CFT which arises from a sigma model
with target space $M_4^N/S_N$. Here $S_N$ is the symmetric group, so the target
space is an orbifold with this symmetry group. In \ref\lmtwo{O.~Lunin and
S.~D.~Mathur, {\it in preparation}} this computation is extended to the
supersymmetric CFT which is expected to correspond to the D1-D5 system at the
orbifold point \ref\martinec{F.~Larsen and E.~Martinec, ``U(1) 
charges and moduli
in the D1-D5 system,'' JHEP{\bf 9906}, 019 (1999) [hep-th/9905064].}.  The
latter computation also allows us to compute correlation functions in 
the Ramond
sector of the CFT, thus giving a `mini black hole S-matrix'.  By the 
observation of the
preceeding paragraph, for the maximally spinning D1-D5 system,
  if we wish to replace the physics of the AdS region with an
effective
  boundary CFT
then this CFT will be in the NS ground state.
It would be
interesting to compare correlation functions in the CFT with 
calculations in the dual string
theory: it is not clear which quantities will be protected against 
change when we
move from the orbifold point to the point in moduli space where the 
supergravity
description is good.

The general issue of whether we can get nonsingular metrics from 
branes  is very
interesting. The 6-brane of type IIA theory lifts to a nonsingular 
brane in 11-D
(the geometry shares some features with the metric \lonep\ at
$\gamma_1=1, \gamma_2=0$).
Also,  we can dualize any brane to a zero brane, which can be 
regarded as just a
gravitational wave in M-theory.  The geometry formed by D3-branes is
nonsingular, but the Poincare patch appears to have a natural extension to a
second region behind the horizon, which effectively doubles up the spacetime
\ref\horo{G.~T.~Horowitz and H.~Ooguri, ``Spectrum of large N gauge theory from
supergravity,'' Phys.\ Rev.\ Lett.\ {\bf 80}, 4116 (1998) 
[hep-th/9802116].}. In
this case perhaps one should orbifold across the horizon, and thus 
keep only one
copy of asymptotic infinity; we would get additional states at the 
orbifold plane
which could be excited in the process of absorption by the branes. 
The metrics of
\bal \mm\ provide new interesting examples where the `matter' representing the
branes has effectively been completely represented by the metric and 
gauge fields
produced by the branes.

\bigskip

{\bf Acknowledgements}
\bigskip

  I am grateful to V. Balasubramanian, S. Das, A. Jevicki and O. Lunin 
for helpful
discussions. This work  is supported in part by DOE grant no. 
DE-FG02-91ER40690.

\appendix{A}{Vector fields on the sphere}

Consider the sphere $S^3$.  We will need to study vector fields on this sphere,
since supergravity fields like $h_{a\mu}, b_{a\mu}$ have one index on 
this sphere,
and so give rise to vector fields on this sphere. The lowest modes of 
these vector
fields will give rise to the modes of interest to us, so we construct 
them explicitly
below.

The symmery group of this sphere is $SO(4)$, which has the algebra
$$[M_{ab}, M_{cd}]= \delta_{bc} M_{ad} + \delta_{ad} M_{bc} 
-\delta_{ac} M_{bd }-
\delta_{bd} M_{ac}$$ The combinations
\eqn\three{J_1={1\over 2}[M_{12}+M_{34}], ~~J_2={1\over 2}[M_{23}+M_{14}],
~~J_3={1\over 2}[M_{13}+M_{42}]} form an $SU(2)$, as do the combinations
\eqn\threep{K_1={1\over 2}[M_{12}-M_{34}], ~~K_2={1\over 2}[M_{23}-M_{14}],
~~K_3={1\over 2}[M_{13}-M_{42}]} Thus we observe the decomposition
$SO(4)=SU(2)\times SU(2)$:
\eqn\four{[J_i, J_j]=\epsilon_{ijk} J_k, ~~~[K_i, K_j]=\epsilon_{ijk} 
K_k, ~~~[J_i, K_j]=0}

To find a representation of this algebra on vector fields on $S^3$, 
consider the unit
sphere
\eqn\one{x_1^2+x_2^2+x_3^2+x_4^2=1}

We can find vector fields on the sphere that  form a  representation of  the
generators $M_{ab}$, by writing $M_{ab}$ as matrices:
\eqn\onep{(M_{ab})_{a'b'}=\delta 
_{aa'}\delta_{bb'}-\delta_{ab'}\delta_{ba'}} Under a
generator $\epsilon M_{ab}$ we get the differmorphism of $S^3$
\eqn\onepp{\delta X_{a'} = \epsilon (M_{ab})_{a'b'}X_{b'}\equiv 
\epsilon {1\over
2}V_{a'} } (Thus these vector fields are the tangent vectors arising from the
nonlinear action of the rotation group on the sphere. They are 
Killing vector fields
for $S^3$.)

The following are three vector fields  arising from the combinations 
$J_i$: -- we list
their components at a point
$(x_1, x_2, x_3, x_4)$:
\eqn\two{ V^{(1)}=\{x2, -x1, x4, -x3\}, ~~ V^{(2)}=\{x4, x3, -x2, -x1\}, ~~
V^{(3)}=\{x3, -x4, -x1, x2\}} These fields have unit norm everywhere, and are
orthogonal to each other at each point of the sphere.

The fields arising from the combinations $K_i$ are
\eqn\twop{ W^{(1)}=\{x2,- x1, -x4, x3\}, ~~ W^{(2)}=\{-x4, x3, -x2, x1\}, ~~
W^{(3)}=\{x3, x4, -x1,- x2\}}

The fields $V^{(i)}$ give a representation $(3,1)$ and the fields 
$W^{(i)}$ give a
representation $(1,3)$ under the
$SU(2)\times SU(2)$. Each set of fields give a representation $3$ under the
diagonal $SU(2)$ subgroup $J_i+K_i$.

\appendix{B}{Symmetries of anti-de-Sitter space}

We can obtain the anti-de-Sitter space $AdS_3$ with unit radius as the surface
\eqn\qone{-x_1^2-x_2^2+x_3^2+x_4^2=-1} The symmetry group is
$SO(2,2)=SU(1,1)\times SU(1,1)$.  Let
\eqn\qtwo{\eta_{ab}=\eta^{ab}={\rm diag}\{ 1,1,-1,-1\} } be a metric 
that will be
used to raise and lower indices. The generators of the algebra can be 
written as the
matrices
\eqn\qel{(\tilde 
M_{ab})^{a'}_{b'}=\delta^{a'}_a\eta_{bb'}-\eta_{ab'}\delta^{a'}_b}
which satisfy the algebra
\eqn\qthree{[\tilde M_{ab}, \tilde M_{cd}]= \eta_{bc} \tilde M_{ad} + 
\eta_{ad} \tilde
M_{bc} -\eta_{ac} \tilde M_{bd }- \eta_{bd}
\tilde M_{ac}}

  The generator $\epsilon \tilde M_{ab}$ acts on $AdS_3$ as
\eqn\qtw{\delta X^{a'} = \epsilon(\tilde M_{ab})^{a'}_{b'}X^{b'}\equiv \epsilon
{1\over 2}V^{a'} }

The combinations
\eqn\qthree{\tilde J_1={1\over 2}[\tilde M_{12}-\tilde M_{34}], 
~~\tilde J_2={1\over
2}[\tilde M_{23}+\tilde M_{14}], ~~\tilde J_3={1\over 2}[\tilde 
M_{13}+\tilde M_{42}]}
  give
\eqn\qfour{[\tilde J_1, \tilde J_2]=\tilde J_3, ~~~[\tilde J_2, 
\tilde J_3]=-\tilde J_1,
~~~[\tilde J_3,\tilde J_1]=\tilde J_2 } and the combinations
\eqn\qthree{\tilde K_1={1\over 2}[\tilde M_{12}+\tilde M_{34}], ~~\tilde
K_2={1\over 2}[\tilde M_{23}-\tilde M_{14}], ~~\tilde K_3={1\over 2}[\tilde
M_{13}-\tilde M_{42}]}  also give the same algebra
\eqn\qfour{[\tilde K_1, \tilde K_2]=\tilde K_3, ~~~[\tilde K_2, 
\tilde K_3]=-\tilde K_1,
~~~[\tilde K_3,\tilde K_1]=\tilde K_2 }

Now we note that the combinations
\eqn\qsix{L_0=i\tilde J_1, ~~~L_1=i\tilde J_2+\tilde J_3, 
~~~L_{-1}=i\tilde J_2-\tilde
J_3}  form the anomaly free part of the Virasoro algebra
\eqn\qseven{[L_0, L_1]=-L_1, ~~~[L_0, L_{-1}]=L_{-1}, ~~~[L_1, 
L_{-1}]=2 L_0} (We
have $L_i=L_{-i}^\dagger$ since the $J_i$ are anti-Hermitian.) We get 
a similar copy
of the Virasoro algebra from the generators $\tilde K_i$.

Let us now choose convenient coordinates on anti-de-Sitter space. In 
the relation
\qone\ write
\eqn\qeight{\eqalign{x_3^2+x_4^2&=r^2, ~~~x_3=r\cos\phi, ~~~x_4=r\sin\phi\cr
x_1^2+x_2^2&=1+r^2, ~~~x_1=\sqrt{1+r^2} \cos t, ~~~x_2=\sqrt{1+r^2}\sin t\cr }}
Then we get on anti-de-Sitter space the metric
\eqn\qnine{ds^2=-(1+r^2)dt^2+{dr^2\over 1+r^2} + r^2d\phi^2} We take the
covering space of this space, so that $-\infty<t<\infty$ (we still have
$0\le\phi<2\pi$).

If we write $r=\sinh\rho$ we get
\eqn\qten{ds^2= -\cosh^2\rho dt^2+d\rho^2+\sinh^2\rho d\phi^2}

Writing the action of the Virasoro generators \qseven\ in terms of 
their action of
functions (an action $X^i\rightarrow X^i+\epsilon V^i$ gives rise to 
an  operation
$-\epsilon V^i\partial_i$) we get
\eqn\qeleven{L_0=i\partial_u}
\eqn\qtw{L_1=i e^{iu}(\rm {Coth}(2\rho)\partial_u -{1\over \rm
{Sinh}(2\rho)}\partial_v -{i\over 2}\partial_\rho)}
\eqn\qtw{L_{-1}=i e^{-iu}(\rm {Coth}(2\rho)\partial_u -{1\over \rm
{Sinh}(2\rho)}\partial_v +{i\over 2}\partial_\rho)} where $u=t+\phi, v=t-\phi$.

\appendix{C}{The $N=4$ algebra and spectral flow}

The theory has a left moving and a right moving chiral algebra, which 
are each of
the form
\eqn\wone{[L_m, L_n]=(m-n) L_{m+n}+{c\over 12} m(m^2-1) \delta_{m+n,0}}
\eqn\wtwo{[J^i_m, J^j_n]=i\epsilon^{ijk}L_{m+n}+{c\over 12} m \delta_{m+n,0},
~~~[L_m, J_n^i]=-nJ^i_{m+n}}
\eqn\wthree{\{G_r^a , \bar G_{b,s}\}=2\delta^a_b L_{r+s}-2(r-s)(\sigma^i)^a{}_b
J^i_{r+s}+{c\over 12} (4 r^2-1)\delta^a_b
\delta_{r+s,0}}
\eqn\wfour{\{G^a_r, G^b_s\}=0, ~~~\{\bar G_{a,r}, \bar G_{b,s}\}=0 }
\eqn\wfive{[L_m, G^a_r]=({m\over 2}-r)G^a_{m+r}, ~~~ [L_m, \bar
G_{a,r}]=({m\over 2}-r)\bar G_{a, m+r}}
\eqn\wsix{[J^i_m, G^a_r]=-{1\over 2}(\sigma^i)^a{}_b G^b_{m+r}, ~~~[J^i_m, \bar
G_{a,r}]={1\over 2}\bar G_{b, m+r}(\sigma^i)^b{}_a}

Under a spectral flow by  parameter $\alpha$ we get
\eqn\wseven{h'=h-\alpha j_3+{c\alpha^2\over 24}}
\eqn\weight{j_3'=j_3-{c\alpha\over 12}} Thus the NS vacuum with $h=0, j_3=0$
flows for $\alpha=1$ to $h'={c\over 24}, j_3'=-{c\over 12}$. This value of $h'$
correspods to a R sector ground state.

While the spectral flow is formally just an automorphism of the algebra, we can
look at the following physical situation. We increase the value of 
the flat connection
starting from zero,  and follow any chosen state of the system.  Focusing for
simpilicity on a U(1) in the gauge group, we note that $\int A 
dx=4\pi$ the theory
returns to what it was at
$A=0$ ( at
$\int A dx=2\pi$ the fermions are antiperiodic even though the bosons have
returned to their original Lagrangian). But the state that we were 
following does
not return to the same state, and thus we get a map from states of 
the system to
states of the same system. This map causes the changes in quantum numbers
listed above under the spectral flow.

  The D1-D5 system has $c=6N=6Q_1Q_5$.  The length scale given by the radius  of
the space
$AdS_3$ and the sphere
$S^3$ is
$l=\sqrt{\alpha'}g_6^{1\over 2}N^{1\over 4}$.  Here
$g_6$ is the 6-dimensional string coupling constant, and is given by
$g_6=g/\sqrt{v}$, where $g$ is the 10-d string coupling constant and the volume
of the compact 4-d space (K3 or
$T^4$) is
$(2\pi)^4 \alpha'^2 v$.

\appendix{D}{Solving the first order equations for a 1-form}

In this appendix we look at the equations \rsevent\ that arise from the coupled
equations for the graviton and the 2-form field. Let us look at a 
more general set of
equations of the form
$$*dW+QW=0$$ where $Q$ is a constant.

Let
$$u=t+\phi, ~~v=t-\phi, ~~~~t={1\over 2} (u+v), ~~\phi={1\over 2} (u-v)$$ Then
$$\epsilon_{uvr}=-{1\over 2}\epsilon_{\hat t\hat\phi \hat r} r$$ We let
$\epsilon_{\hat t\hat\phi \hat r}=1$.

Since the background has translational invariance in $t, \phi$ we can separate
harmonics in these variables. We look at an ansatz of the form
$$W_u=e^{-i\alpha u} f(r)$$
$$W_r=e^{-i\alpha u} h(r)$$
$$W_v=e^{-i\alpha u} q(r)$$ (The parameter $\alpha$ here does not have any
relation to the $\alpha$ used earlier in the paper to denote the `twist' of the
spacetime; we are at present looking at perturbations around $AdS_3\times
S^3\times M_4$ without any twist.) Then we have
$$F_{ur}=W_{r,u}-W_{u,r}=[-i\alpha h - f']e^{-i\alpha u}$$
$$F_{vr}=W_{r,v}-W_{v,r}=[ - q']e^{-i\alpha u}$$
$$F_{uv}=W_{v,u}-W_{u,v}=[-i\alpha q ]e^{-i\alpha u}$$ This gives
\eqn\done{(*F)_u=-{1\over 2r}[ (1+2r^2)(i\alpha h + f')-q']e^{-i\alpha u}}
\eqn\dtwo{(*F)_v=-{1\over2r}[(i\alpha h + f')-(1+2r^2)q']e^{-i\alpha u}}
\eqn\dthree{(*F)_r=-{2i\alpha\over r(1+r^2)} qe^{-i\alpha u}}

 From \dthree\ we get
\eqn\xone{h={2i\alpha\over Q r(1+r^2)} q} From \done , \dtwo\ we get
\eqn\xtwo{f={2(r+r^3)q'\over Q}+(1+2r^2) q} Putting this in \dtwo\ we get
\eqn\xthree{r(1+r^2)[r(1+r^2)q']' +[Q(2-Q)r^2(1+r^2)-{\alpha^2}] q=0} Let us
write
$$Q(2-Q)=P, ~~~r^2=x, ~~~q=({r\over \sqrt{1+r^2}})^\alpha ~z$$ We then get the
equation of hypergeometric type
\eqn\xfour{x(1+x)z_{xx}+(\alpha +1+2x)z_x+{P\over 4}z=0}

For $Q=2$ (which is one of the cases in \rsevent ) we get $P=0$. Then 
we get the
solution
$$q=x^{\alpha/2}(1+x)^{-\alpha/2}[-{1\over \alpha} {(1+x)^\alpha\over
x^\alpha}+Cx+D]$$

\appendix{E}{The wave equation for scalars in the `twisted' geometries}

Consider a massless scalar field in the `twisted' geometry \zone . 
For an ansatz
$$f=e^{-i\omega t}f(r)Y^{l_1,0}(y)$$
  we get the equation for the radial function
\eqn\fone{{1\over r} [ f_{,r} r (1+r^2)]_{,r}+ \omega^2 (1+r^2)^{-1}f-
[l_1(l_1+2)+\alpha^2(j_3+k_3)^2r^{-2}]f=0} Writing
$$r^2=x, ~~~f=r^{-\alpha(j_3+k_3)}(1+r^2)^{-{1\over 2}\omega}z$$ we get the
equation of hypergeometric form
$$x(1+x)z_{xx}+z_x[(1-\alpha(j_3+k_3))  (1+x)+(1-\omega) x]+{P\over 4}z=0$$
where
$$P=(\omega+\alpha(j_3+k_3))(\omega+\alpha(j_3+k_3)-2)-l_1(l_1+2)$$ The
solutions in section 4 with $\omega=l_1+2-\alpha(j_3+k_3)$ have $P=0$,  $z=1$.

If we include a dependence $e^{i m \phi}$ on the angular coordinate 
of $AdS$, then
the equation \fone\  becomes
\eqn\fonep{{1\over r} [ f_{,r} r (1+r^2)]_{,r}+ \omega^2 (1+r^2)^{-1}f-
[l_1(l_1+2)+(\alpha(j_3+k_3)-m)^2r^{-2}]f=0} Thus we just get a  replacement of
$\alpha(j_3+k_3)$ by $\alpha(j_3+k_3)-m$ in the above discussion.

\bigskip
\bigskip

\listrefs
\end